# A Chaos-based Image Encryption Scheme using Chaotic Coupled Map Lattices

Sodeif Ahadpour
Department of Physics, University of Mohaghegh Ardabili, Ardabil, IRAN

Yaser Sadra
Department of Physics, University of Mohaghegh Ardabili, Ardabil, IRAN

**Abstract:** In recent years, information security essential in various arenas like internet communication, multimedia systems, medical imaging, tele-medicine and military communication. However, most of them faced with some problems such as the lack of robustness and security. *In this letter, after reviewing the main points of the chaotic trigonometric maps and the coupled map lattices, we introduce the scheme of chaos-based image encryption based on coupled map lattices. The scheme decreases periodic effect of the ergodic dynamical systems in the chaos-based image encryption. To evaluate the security of the encrypted image of this scheme, the key space analysis, the correlation of two adjacent pixels and differential attack were performed. This scheme tries to improve the problem of failure of encryption such as small key space and level of security.*

## 1. Introduction

In recent years, information security essential in various arenas like internet communication, multimedia systems, medical imaging, tele-medicine and military communication, and so on, leading to an increasing interest in the field of cryptography [1-4]. The cryptographic schemes have proposed some novel and efficient methods to develop secure of the image encryption [5-7]. The chaos-based encryption schemes are composed of two steps: chaotic confusion and pixel diffusion. In the chaotic confusion stage, a combination of the chaotic maps is used to realize the confusion of all pixels. The parameters of the chaotic maps are used for the confusion key. In the pixel diffusion stage, a plain image permutes or the value of each pixel changes one by one with using of the chaotic confusion stage. The parameters of the diffusion function are used for the diffusion key. The most of schemes be faced with some problems such as the lack of robustness and security. In order to improve the problems, we proposed a chaos-based image encryption scheme using chaotic coupled map lattices. Here, we use of the chaotic trigonometric maps for increase the space of the confusion key. This paper is arranged as follows. In section 2, the chaotic confusion stage is discussed. In section 3, we propose a chaos-based image encryption scheme based on the coupled map lattices. In section 4, the analysis of security of the proposed encryption scheme is discussed and finally, in Section 5, we conclude the paper.

## 2. The Chaotic Confusion Stage

In this section, we proposed a combination of the chaotic trigonometric and the coupled map lattices as the chaotic confusion stage. This method increases the space of the confusion key, that be caused the development of robustness and security.

### 2.1. The Chaotic Trigonometric

We first review one-parameter chaotic maps which can be used in the construction of chaotic trigonometric maps. The one-parameter chaotic maps [8] are defined as the ratio of polynomials of degree N:

$$\phi_N^1(x,a) = (1+(-1)^N {}_2F_1(-N,N,\frac{1}{2},x)) \times \frac{a^2}{(a^2+1)+(a^2-1)(-1)^N {}_2F_1(-N,N,\frac{1}{2},x)}$$

$$= \frac{a^2(T_N(x^{\frac{1}{2}}))^2}{1+(a^2-1)(T_N(x^{\frac{1}{2}}))^2}$$

and

$$\phi_N^2(x,a) = (1-(-1)^N {}_2F_1(-N,N,\frac{1}{2},(1-x))) \times \frac{a^2}{(a^2+1)-(a^2-1)(-1)^N {}_2F_1(-N,N,\frac{1}{2},x)}$$

$$= \frac{a^2(U_N((1-x)^{\frac{1}{2}}))^2}{1+(a^2-1)(U_N((1-x)^{\frac{1}{2}})^2)}$$

where N is an integer greater than one. Also,

$$_2F_1(-N,N,\frac{1}{2},x)=(-1)^N\cos(2N\arccos(x^{\frac{1}{2}}))$$
$$=(-1)^N T_{2N}(x^{\frac{1}{2}})$$

is the hypergeometric polynomials of degree N and $T_N(U_n(x))$ are chebyshev polynomials of type I (typeII), respectively. The chaotic trigonometric maps are their conjugate maps which are defined as:

$$\begin{cases}\tilde{\phi}_N^1(x,a)=h\circ\phi_N^1(x,a)\circ h^{-1}\\=\frac{1}{a^2}\tan^2(N\arctan(x^{\frac{1}{2}})),\\ \tilde{\phi}_N^2(x,a)=h\circ\phi_N^2(x,a)\circ h^{-1}\\=\frac{1}{a^2}\cot^2(N\arctan(x^{-\frac{1}{2}})).\end{cases} \quad (1)$$

Conjugacy means that invertible map $h(x)=\frac{1-x}{x}$ maps I =[0, 1] into [0,∞) [8]. In order to simplify the calculation in this paper, we denote the chaotic trigonometric maps ($\tilde{\phi}_N^1(x,a),\tilde{\phi}_N^2(x,a)$) with $f_1(x,a), f_2(x,a)$ respectively. Therefore, the chaotic trigonometric maps are as follows:

$$\begin{cases}f_1(x_n,a_1)=\frac{1}{a_1^2}\tan^2(N_1\arctan(x_{n-1}^{\frac{1}{2}})),\\ f_2(x_n,a_2)=\frac{1}{a_2^2}\cot^2(N_2\arctan(x_{n-1}^{-\frac{1}{2}})).\end{cases}$$

## 2.2. The Chaotic Coupled Map Lattices

The coupled map lattices are arrays of states whose values are continuous, usually within the unit interval, or discrete space and time. The coupled map is as a two-dimensional dynamical map which is defined as:

$$f(X,Y)=\begin{cases}X=f(x,y)\\Y=f(y,x)\end{cases}$$

In this paper, we propose the chaotic coupled map lattices as generic symmetric non-linearly coupled maps which are the two-dimensional dynamical chaotic maps as following:

$$f_{coupled}(X,Y)=\begin{cases}X=(1-\varepsilon)f_1(x)+\varepsilon f_2(y)\\Y=(1-\varepsilon)f_1(y)+\varepsilon f_2(x)\end{cases} \quad (2)$$

where, $\varepsilon$ is the strength of the coupling, and the functions $f_1$ and $f_2$ are the chaotic trigonometric maps.

## 3. Encryption scheme based on Chaotic Coupled Map Lattices

In the pixel diffusion stage, we proposed an encrypted scheme based on the coupled map lattices. In the proposed scheme, the composite of the chaotic coupled map lattices are employed to achieve the goal of image encryption. To consider a gray scale image with the size of m×n ($I_{m\times n}$). The gray scale image $I_{m\times n}$ is transformed into the matrix $I_{(m\times n)\times 1}$. Then, using the results of iteration of the chaotic logistic map [9]
$x_{n+1}=rx_n(1-x_n)$ ( $x_n\in(0,1)$ and $r_x\in(3.99996,4)$),
we generate the strength of the coupling $\varepsilon$ ($\varepsilon=x_{n+1}$).
Now, using the chaotic coupled map lattices (Eq. 2), the encryption shceme is defined as

$$\begin{cases}E_i=\left[(X_i\times 10^{14})\mod(m\times n)\right]\oplus I_{i\times 1}\\ F_i=\left[(Y_i\times 10^{14})\mod(m\times n)\right]\oplus I_{i\times 1}\end{cases}$$

Therefore,
$$C_i=E_i\oplus F_i \quad (3)$$

where C and $\oplus$ are the matrix of the encrypted image and the bitwise XOR operator, respectively. Note that $X_i, Y_i$ are, in fact, the results of the iteration of the chaotic coupled map lattices. The decryption process is almost the same as the encryption but with reverse steps.

An indexed image of an 'Albert Einstein' sized $256\times 256$ (see Fig. 1(a)) is used as a plain image and the encrypted image is shown in Fig. 1(b). The grey scale histograms are given in Fig. 2. The Fig. 2(b) shows uniformity in distribution of grey scale of the encrypted image. In addition, the average pixel intensity for plain image is 98.92, and for encrypted image is 127.60, respectively.

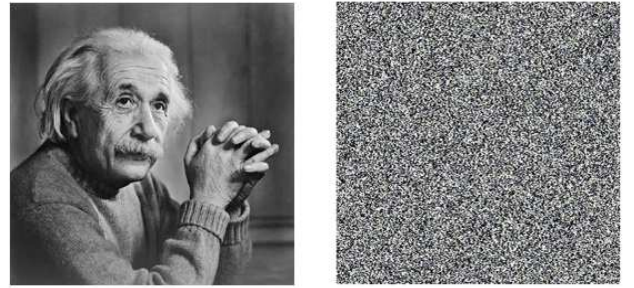

Figure 1. (a) Plain image; (b) Encrypted image.

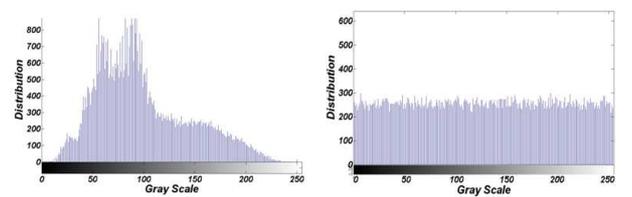

Figure 2. Histograms of images.

## 4. Analysis of security of the proposed encryption scheme

The Security is a major intransitive of a cryptosystem. Here, a complete analysis is made on the security of the cryptosystem. We have tried to explain that this cipher image is sufficiently secure against various cryptographical attacks, as shown below:

## 4.1. Key space analysis

In the proposed scheme, the confusion stage and diffusion stage are applied respectively. Thus, the key space of the encryption is the multiplication between the confusion key and the diffusion key, i.e. $S = S_1 + S_2$ where $S_1$, $S_2$ and $S$ are the confusion key, the diffusion key and the key space, respectively. On the other hand, the key space size is the total number of different keys that can be used in the encryption [10,14]. Security issue is the size of the key space. If it is not large enough, the attackers may guess the image with brute-force attack. If the precision is $10^{-14}$, the size of key space for initial conditions and control parameters of the proposed scheme is more than $2^{302}$. This size is large enough to defeat brute-force by any super computer today.

## 4.2. Correlation Coefficient analysis

The statistical analysis has been performed on the encrypted image. This is shown by a test of the correlation between two adjacent pixels in plain image and encrypted image. We randomly select 2000 pairs of two-adjacent pixels (in vertical, horizontal, and diagonal direction) from plain images and ciphered images, and calculate the correlation coefficients, respectively by using the following two equations (see Table 1 and Fig. 3) [10,11]:

$$Cov(x,y) = \frac{1}{N}\sum_{i=1}^{N}(x_i - E(x))(y_i - E(y)),$$

$$r_{xy} = \frac{Cov(x,y)}{(D(x))^{\frac{1}{2}}(D(y))^{\frac{1}{2}}}$$

where

$$E(x) = \frac{1}{N}\sum_{i=1}^{N}(x_i), \quad D(y) = \frac{1}{N}\sum_{i=1}^{N}(x_i - E(x))^2.$$

where, E(x) is the estimation of mathematical expectations of x, D(x) is the estimation of variance of x, and Cov(x,y) is the estimation of covariance between x and y, where x and y are grey scale values of two adjacent pixels in the image.

Table 1: Correlation coefficients of two adjacent pixels in the plain image and the encrypted image

| Direction | **Plain Image** | Encrypted Image |
|---|---|---|
| **Horizontal** | 0.9341 | 0.0014 |
| **Vertical** | 0.9634 | 0.0036 |
| **Diagonal** | 0.9402 | 0.0027 |

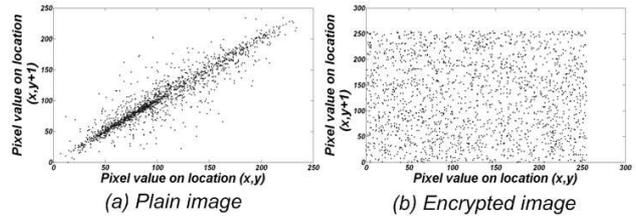

(a) Plain image      (b) Encrypted image

Figure 3. Correlation distributions of two horizontally adjacent pixels in the plain image and the encrypted image.

## 4.3. Differential attack

Attackers try to find out a relationship between the plain image and the encrypted image, by studying how differences in an input can affect the resultant difference at the output in an attempt to derive the key [12,15]. Trying to make a slight change such as modifying one pixel of the plain image, attacker observes the change of the encrypted image [12]. Because of the existence of the diffusion in the proposed cryptosystem, the encrypted image is so sensitive to the plain image that even a one-pixel change in the plain image leads to a completely different encrypted image. Diffusion refers, in fact, to rearrange or spread out the bits in the message. So, any redundancy in the plain image is spread out over the encrypted image [13,10]. In order to demonstrate influence of one pixel change on the whole encrypted image by the proposed scheme, two common measures are used:

Number of Pixels Change Rate (NPCR) stands for the number of pixels change rate while, one pixel of plain image is changed. Unified Average Changing Intensity (UACI) measures the average intensity of differences between the plain image and ciphered image. The NPCR and The UACI, are used to test the influence of one pixel change on the whole image encrypted by the proposed scheme and can be defined as following:

$$NPCR = \frac{\sum_{i,j} D(i,j)}{W \times H} \times 100\%$$

$$UACI = \frac{1}{W \times H}\left[\sum_{i,j} \frac{C_1(i,j) - C_2(i,j)}{255}\right] \times 100\%$$

where $W$ and $H$ are the width and height of $C_1$ or $C_2$. $C_1$ and $C_2$ are two ciphered images, whose corresponding original images have only one pixel difference and also have the same size. The $C_1(i,j)$ and $C_2(i,j)$ are grey-scale values of the pixels at grid (i,j). The D(i,j) determined by $C_1(i,j)$ and $C_2(i,j)$. If $C_1(i,j) = C_2(i,j)$, then, D(i, j) = 1; otherwise, D(i, j) = 0. We have done some tests on the proposed scheme (256 grey scale image of size $256 \times 256$) to find out the extent of change produced by one pixel change in the plain image. We have obtained

NPCR=0.25% and UACI=0.19%. The results demonstrate that the proposed scheme can survive differential attack.

## 5. Conclusion

We have proposed a chaos-based encryption scheme based on *coupled map lattices*. The security of the encrypted image of this scheme is evaluated by the key space analysis, the correlation of two adjacent pixels and differential attack. The distribution of the encrypted image is very close to the uniform distribution, which can well protect the information of the image to withstand the statistical attack. We suggest that this encryption scheme is suitable for applications like internet image encryption and secure transmission of confidential information in the internet.

## References

[1] Ahadpour S., Sadra Y., "Randomness criteria in binary visibility graph and complex network perspective" Information Sciences, 197, pp. 161–176, 2012.

[2] Guo X., Zhang J., "Secure group key agreement protocol based on chaotic Hash," Information Sciences, 1180, pp. 4069–4074, 2010.

[3] Liu S., Long Y., Chen K., "Key updating technique in identity-based encryption," Information Sciences, 181, pp.2436–2440, 2011.

[4] Weng J., Yao G., Deng R. H., Chen M., Li X., "Cryptanalysis of a certificateless signcryption scheme in the standard model," Information Sciences, 181, pp. 661–667, 2012.

[5] Pareek N. K., Patidar V., Sud K. K., "Image encryption using chaotic logistic map," Image and Vision Computing, 24, pp. 926–934, 2006.

[6] Ahadpour S., Sadra Y., ArastehFrad Z., "A Novel Chaotic Image Encryption using Generalized Threshold Function" International Journal of Computer Applications, 42(18), pp. 25-31, 2012.

[7] Ahadpour S., Sadra Y., ArastehFrad Z., "A Novel Chaotic Encryption Scheme based on Pseudorandom Bit Padding" IJCSI International Journal of Computer Science Issues, 9(1), 449-456, 2012.

[8] Jafarizadeh M.A., Foroutan M., Ahadpour S., "Hierarchy of rational order families of chaotic maps with an invariant measure," Pramana-journal of physics, 67, pp. 1073-1086, 2006.

[9] Kanso A., Smaoui N., "Logistic chaotic maps for binary numbers generations," Chaos, Solitons and Fractals, 40, pp. 2557-2568, 2009.

[10] Behnia S., Akhshani A., Ahadpour S., Mahmodi H., Akhavan A., " A fast chaotic encryptionscheme based on piece wise nonlinear chaotic maps," Physics Letters A, 366, pp. 391–396, 2007.

[11] Chen G., Mao Y., Chui C.K., "A symmetric image encryption scheme based on 3D chaotic cat maps," Chaos Solitons Fractals, 21, pp. 749-761, 2004.

[12] Maniyath Sh. R. and Supriya M., "An Uncompressed Image Encryption Algorithm Based on DNA Sequences," Computer Science and Information Technology, pp. 258–270, 2011.

[13] Smaoui N., Kostelich E., "Using chaos to shadow the quadratic map for all time," Int J Comput Math., 70 pp. 117-129, 1998.

[14] Zhou Q., Wong K.W., Liao X., Xiang T., Hu Y., "Parallel image encryption algorithm based on discretized chaotic map," Chaos Solitons Fractals, 38 pp. 1081-1092, 2008.

[15] Kandar S., Maiti A., Dhara B. C., "Visual Cryptography Scheme for Color Image Using Random Number with Enveloping by Digital Watermarking," IJCSI International Journal of Computer Science Issues, 8, pp. 543-549, 2011.